\documentclass[11pt,aps,preprint,nofootinbib]{revtex4}
\usepackage[active]{srcltx}
\usepackage[utf8]{inputenc}
\usepackage{latexsym}
\usepackage{amsmath}
\usepackage{graphicx}

\begin{document}

\title{Aspects of Emergent Symmetries}

\author{Pedro R. S. Gomes}
\email{pedrogomes@uel.br, pedrorsg.gomes@gmail.com}
\affiliation{Departamento de F\'isica, Universidade Estadual de Londrina, \\
Caixa Postal 10011, 86057-970, Londrina, PR, Brasil}
\affiliation{Instituto de F\'\i sica, Universidade de S\~ao Paulo\\
Caixa Postal 66318, 05314-970, S\~ao Paulo, SP, Brazil}

%%%%%%%%%%%%%%%%%%%%%%%%%%%%%%%%%%%%%%%
\begin{abstract}

These are intended to be review notes on emergent symmetries, i.e., symmetries
which manifest themselves in specific sectors of energy in many systems.
The emphasis is on the physical aspects rather than computation methods.
We include some background material and go through more recent problems
in field theory, statistical mechanics and condensed matter. 
These problems illustrate how some important symmetries, such as Lorentz invariance and supersymmetry, usually believed to be
fundamental, can arise naturally in low-energy regimes of systems involving a large number of degrees of freedom.
The aim is to discuss how these examples could help us to face other complex and fundamental problems.

\end{abstract}
\maketitle

%%%%%%%%%%%%%%%%%%%%%%%%%%%%%%%%%%%%%%%%%%%%

\section{Introduction}\label{Int}

Two of the most notable facts of physics that actually enable us to make progress in understanding nature
are {\it scale-dependence} and the {\it decoupling} of different scales. 
In a given system we usually observe dynamics at different levels because it
possesses distinct relevant degrees of freedom acting at different scales.
These ideas are clearly illustrated by a fluid.
We need different theories to describe it as we change the scale length $l$ \cite{Gross}:
\begin{eqnarray}
l&\sim&1\text{cm}  \Rightarrow \text{Navier-Stokes theory}\nonumber\\
l&\sim&10^{-5}\text{cm} \Rightarrow \text{granular theory}\nonumber\\
l&\sim& 10^{-8}\text{cm} \Rightarrow \text{atomic theory}\nonumber\\
l&\sim& 10^{-13}\text{cm} \Rightarrow \text{nuclear physics}\nonumber\\
l&\sim& 10^{-13} - 10^{-18}\text{cm} \Rightarrow \text{quantum chromodynamics} \nonumber\\
l&\sim& 10^{-33}\text{cm} \Rightarrow \text{new physics (string theory, discrete spacetime, etc)}.\nonumber
\end{eqnarray}
The {\it scale-dependence} and the {\it decoupling} are evident.
To understand the macroscopic dynamics of the fluid, described by the Navier-Stokes equation,  it is not necessary to
know string theory!

These features naturally lead us to the idea that some properties, in particular,
symmetries can emerge or disappear as we probe the system in different energy scales.
The main goal of this work is to review some examples where the emergence of
symmetries takes place, and how this can help us to face other complex and fundamental problems.
This is a quite wide subject. Naturally the
problems discussed here reflect personal interests and are not intended to be exhaustive.
We choose particular examples that illustrate some special features we want to stress, including 
recent developments.
A comprehensive account on these subjects can be found in the outstanding reference \cite{Nielsen}.

A very simple example of an emergent symmetry is provided by a two-dimensional flat surface, as the surface of a table.
Consider the motion of a probe-particle on that table.
From our eyes perspective, which has a low power of resolution (low-energy),
the surface seems to be perfectly flat, i.e., invariant under translations.  The motion of the particle on this surface
is invariant under translations.
Now, suppose that we are equipped with a microscope with a higher power of resolution.
We can explore this system in more details with more energy. In this case, we realize that the
system is no longer translational invariant due to the roughness of the surface.
To describe the motion of the probe-particle we need to take into account the presence of roughness, which
breaks the translational invariance.
So when we go from high to low energies we see that the translational invariance is an emergent symmetry.

We can think of two types of symmetries:  {\it fundamental} and {\it emergent}. We say that
a symmetry is {\it fundamental} when it is a symmetry of the whole spectrum of energy.
This is the opposite of an {\it emergent} symmetry, which only manifests itself in specific sectors.
The separation is tenuous once it depends on our capacity in exploring physics through several scales. 
Most of physical theories consider Lorentz symmetry as fundamental, but the possibility
of Lorentz violation in certain situations gives rise to interesting developments. In this sense,
we can think of Lorentz invariance as an emergent symmetry. 

These are intended to be review notes with main focus on the physical aspects rather than the computation methods.
The basic framework to perform this type of investigation involves
the renormalization group and the construction of effective field theories.
The renormalization group essentially tells us how a certain system reacts under scale transformations, while
the effective field theory method enable us to obtain low-energy description of the systems based on the
identification of the relevant degrees of freedom and the underlying symmetries.
We proceed to our discussion by examining several interesting problems in field theory, statistical mechanics and condensed matter.

The work is organized as follows. We start by discussing some basic ingredients of the renormalization group and
of the effective field theory methods in Secs. \ref{RG} and \ref{EFT}, respectively.
In Sec. \ref{Lorentz} we discuss a first example of an emergent symmetry, Lorentz symmetry, which will be instructive to
classify the symmetry-breaking type operators in Sec. \ref{SoftHard}, where are also discussed further examples involving gauge and supersymmetry.
Next we pay particular attention to scale as well as conformal invariance in Secs. \ref{CP} and \ref{CI},  since they are
ubiquitous in phase transitions of condensed matter systems, in particular in two dimensions.
In Sec. \ref{AS} we discuss examples of accidental symmetries, occurring due to some coincident combination
of degrees of freedom. Sec. \ref{Discrete} deals with the ultraviolet divergences in field theory, in particular in gravitation
and how this hints for a discrete nature of the spacetime. We also make some comments from the string theory perspective.
We close the work with some final considerations in Sec. \ref{FC}.

%%%%%%%%%%%%%%%%%%%%%%%%%%%%%%%%%%%%%%%%%%%%%
\section{Few words about Renormalization Group}\label{RG}

The basic framework to understand the behavior of a system under a scale changing is the renormalization group.
We will proceed to our discussion bearing in mind a field theory but of course this is a very general procedure and could, for example,
be equally described in terms of spin systems defined on lattices.
From that perspective, the system is given by a Lagrangian depending on a field $\varphi$
(or a set of fields but we will consider just one for simplicity), its derivatives
and on a set of parameters $\{g_i\}$ (mass, coupling constants, etc)
\begin{equation}
\mathcal{L}=\mathcal{L}(\varphi,\partial\varphi,\partial^2\varphi,...,\{g_i\}).
\label{2.1}
\end{equation}
A basic aspect of the renormalization group is to compare physical phenomena at different scales.
To this end we need to introduce an auxiliary (arbitrary) scale, denoted by $\mu$, which is 
associated to the typical energy we are working (in the next section it will be clear how this can be done).
This is indeed an inherent property of the renormalization program.
It follows that any physical quantity, $\mathcal{F}$, will be a function of these parameters:
\begin{equation}
\mathcal{F}=\mathcal{F}(\{g_i\},\mu).
\label{2.2}
\end{equation}
But does it mean that the physical quantities depend on an arbitrary parameter?
The only possibility for making sense of this is to assume that the parameters $\{g_i\}$ are functions
of $\mu$ too,
\begin{equation}
g_i\rightarrow g_i(\mu).
\label{2.3}
\end{equation}
They become then effective parameters, also called running couplings,
such that physical quantities should be invariant under changing of $\mu$.
This is expressed by the condition
\begin{equation}
\mu\frac{d }{d\mu}\mathcal{F}(\{g_i(\mu)\},\mu)=0,
\label{2.4}
\end{equation}
or,
\begin{equation}
\left(\mu\frac{\partial}{\partial\mu}+\sum_i \beta_i \frac{\partial}{\partial g_i}\right)\mathcal{F}(\{g_i(\mu)\},\mu)=0,
\label{2.5}
\end{equation}
which is usually called a renormalization group equation. Other objects of the quantum theory such as Green or correlation functions 
satisfy also renormalization group equations. 
The $\beta$-functions are defined as,
\begin{equation}
\beta_i\equiv \mu\frac{\partial g_i}{\partial\mu}.
\label{2.6}
\end{equation}
They dictate how the parameters of theory behave under changing of $\mu$.
By computing the $\beta$-functions we are able to understand how the system reacts
under scale transformations.
Of special interest are the zeros of the $\beta$-functions, characterized by
\begin{equation}
\beta_i(\mu^{\ast})=0.
\label{2.7}
\end{equation}
The values $\mu=\mu^{\ast}$ are fixed points of the renormalization group flow and once they are reached the
corresponding parameters stop to flow. An important question is about the nature of a fixed point, i.e.,
it can be attractive or repulsive. Given a configuration in the parameter space near the fixed point,
the system will flow toward or outward it depending on its nature.

%%%%%%%%%%%%%%%%%%%%%%%%%%%%%%%%%%%%%%%%%%%%
\section{Effective Field Theory for Everything}\label{EFT}

We have discussed in general terms some essential points of the renormalization group.
But what about the Lagrangian $\mathcal{L}$?
We will adopt the effective field theory approach to construct Lagrangians.
This is a very powerful method to deal with systems involving a large number of degrees of freedom.
Roughly speaking, the strategy is try to identify the relevant degrees of freedom of the
low-energy physics,  compatible with a certain set of symmetries,
and then describe the system in terms of a class of operators (terms in the Lagrangian) responsible for
the that dynamics.
This can be systematically done by means of the effective field theory analysis described below,
usually referred as Wilson approach \cite{Wilson}.
We follow the Polchinski 's nice presentation in \cite{Polchinski}.

The starting point of a quantum theory involving a scalar field $\varphi$ is the generating functional
\begin{equation}
Z=\int\mathcal{D}\varphi e^{i S[\varphi]},
\label{3.0}
\end{equation}
where $S[\varphi]=\int d^Dx \mathcal{L}$ is the action and $D$ is the spacetime dimension.
Our strategy is to isolate low-energy physics from high-energy physics.
To this end, we introduce some cutoff scale $\Lambda$, which supposedly do that separation. We are interested in studying
the physics at scales $\mu$ much below $\Lambda$, i.e., $\mu<<\Lambda$.
The field $\varphi$ can be decomposed in two parts corresponding to the high-energy modes, $\varphi_H$, for momenta $k>\Lambda$, and
low-energy modes, $\varphi_L$, for $k<\Lambda$,
\begin{equation}
\varphi=\varphi_H+\varphi_L.
\label{3.1}
\end{equation}
This can be made precise in the Fourier space,
\begin{equation}
\mathcal{D}\varphi \equiv \prod_k d\varphi(k)=\prod_{k<\Lambda} d\varphi_L(k)\prod_{k>\Lambda} d\varphi_H(k).
\label{3.1a}
\end{equation}
The generating functional is then factorized according to
\begin{equation}
Z=\int \mathcal{D}\varphi_H \mathcal{D}\varphi_L e^{iS[\varphi_H,\varphi_L]}.
\label{3.2}
\end{equation}
The integration over high-energy modes can be used to produce
\begin{equation}
Z=\int \mathcal{D}\varphi_L e^{iS_{\Lambda}[\varphi_L]},
\label{3.3}
\end{equation}
with the effective action, $S_{\Lambda}[\varphi_L]$, formally defined by
\begin{equation}
e^{iS_{\Lambda}[\varphi_L]}=\int \mathcal{D}\varphi_H e^{iS[\varphi_H,\varphi_L]},
\label{3.4}
\end{equation}
such that the theory is now defined by the functional (\ref{3.3}) which has no explicit dependence on $\varphi_H$.
In general, the effective action can be expanded in terms of local operators
\begin{equation}
S_{\Lambda}=\int d^Dx \sum_i g_i \mathcal{O}_i,
\label{3.5}
\end{equation}
that are constructed out from the fields $\varphi_L$.
A bit of dimensional analysis is useful to estimate the typical order of the operators in $S_{\Lambda}$.
In the natural unit system, both the speed of light and the Planck constant are settled equal to the unity, $c\equiv 1$ and $\hbar\equiv 1$, such that
we have only mass $[M]$ (= energy) dimension or, equivalently, length $[L]$ (= time) dimension, such that $[M]=-[L]$.
We will write the dimension of all operators in mass units. Let $[\mathcal{O}_i]$ be the dimension of the
operator $\mathcal{O}_i$ in mass units. Because the action is dimensionless, it follows that dimensions of the couplings $g_i$ are
\begin{equation}
[g_i]=D-[\mathcal{O}_i].
\label{3.6}
\end{equation}
For a process with typical energy $\mu$, the contribution of the operator $\mathcal{O}_i$ to the action is of the order
\begin{equation}
\int d^Dx \mathcal{O}_i\sim \mu^{[\mathcal{O}_i]-D}.
\label{3.7}
\end{equation}
To make the cutoff dependence explicit, we introduce the dimensionless coupling constants
$\lambda_i=\Lambda^{[\mathcal{O}_i]-D}g_i$. Thus the typical order of the $i$-th term is
\begin{equation}
\lambda_i\left(\frac{\mu}{\Lambda}\right)^{[\mathcal{O}_i]-D}.
\label{3.8}
\end{equation}
The simple dimensional analysis above provides an important hint. It enable us to identify three classes of operators:
\begin{eqnarray}
&&[\mathcal{O}_i]-D>0\Rightarrow \text{irrelevant or nonrenormalizable}\nonumber\\
&&[\mathcal{O}_i]-D=0\Rightarrow \text{marginal or strictly renormalizable }\nonumber\\
&&[\mathcal{O}_i]-D<0\Rightarrow \text{relevant or super-renormalizable }.
\label{3.9}
\end{eqnarray}
We see that for low energies, $\mu\rightarrow 0$, only marginal (renormalizable) and relevant operators contribute to the action.
Irrelevant operators are suppressed by powers of $\mu/\Lambda$.
That analysis furnishes a systematic way to construct effective field theories even if we do not have much information
about high-energy physics: we just have to  include into the Lagrangian all marginal and relevant operators
compatible with the symmetries of the problem. We have good chances to get a satisfactory description
of the low-energy dynamics of the system.

We have so far discussed some steps of the renormalization group without mentioning a single word about one of the well-known
issues of any theory involving quantized fields, namely the ultraviolet (UV) divergences. They show up as a direct consequence
of the quantization rule 
\begin{equation}
[\varphi({\bf x},t),\Pi({\bf y},t)]=i\delta({\bf x}-{\bf y}),
\end{equation}
where $\Pi\equiv{\partial \mathcal{L}}/\partial (\partial_0\varphi)$ is canonical momentum, that is ill-defined at coincident points.
We want to discuss about this now. This is useful to clarify the meaning of the QFT-nomenclature used in (\ref{3.9}).
Field theories are classified according to their operator content as follows.
A theory involving only marginal and relevant operators is called {\it renormalizable}.
In contrast, a theory that contains at least one irrelevant operator is called {\it nonrenormalizable}.
The difference between them is how the divergences appear in each case.
Roughly speaking, when we start the perturbative calculations we end up with divergences for large
momenta (equivalently, short distances). In renormalizable theories, all divergences can be absorbed into the
parameters of the theory. This procedure corresponds simply to a redefinition (or a renormalization) of the original parameters.
However, this is not possible in the case of nonrenormalizable theories, where there are some divergences that
cannot be absorbed in any parameter of the Lagrangian. We could simply add new operators in order to absorb such divergences.
The problem is that required operators in general induce more and more divergences which in turn require more and more operators to absorb them.
The process never stops.

A remarkable feature of the Wilson approach is that it deals with the UV divergences in a more natural way due to
the process of separation of energy scales. The divergences are tamed by the cutoff $\Lambda$.
If we are interested in low-energy physics, these ultraviolet effects are not important at all.
In a certain sense, the Wilson approach unifies our view of renormalizable and nonrenormalizable theories, unveiling the
intrinsic effective nature in both cases.

%%%%%%%%%%%%%%%%%%%%%%%%%%%%%%%%%%%%%%%%%%%

%%%%%%%%%%%%%%%%%%%%%%%%%%%%%%%%%%%%%%%%%%%
\section{Lorentz Invariant world?}\label{Lorentz}

Now that we have some basic ingredients of the renormalization group and effective field theories,
we are ready to explore some interesting examples. We start by discussing
some issues concerning Lorentz invariance,  where it is realized as an IR fixed point of the renormalization group.

Consider a theory with a single scalar field with a cubic self-interaction given by the Lagrangian,
\begin{equation}
\mathcal{L}=\frac12 \partial_{0}\varphi \partial_{0}\varphi-\frac{b_{\varphi}^2}{2} \partial_{i}\varphi \partial_{i}\varphi-\frac{m^2}{2}\varphi^2
-\frac{\lambda}{3!}\varphi^3.
\label{q1a}
\end{equation}
Latin indices $i,j,...$ assigned to coordinates run only over spatial dimensions while Greek indices $\mu,\nu,...$ run over both space and time dimensions.
As $[x^{\mu}]=-1$ in mass units, it is easy to check that $[\varphi]=(D-2)/2$ and then $[\lambda]=(6-D)/2$, which 
shows that the model is renormalizable in $5+1$ spacetime dimensions.
We want to study the renormalization group
flow of the parameter $b_{\varphi}$, which is defined according to (\ref{2.6}), i.e.,
$\beta_{b_{\varphi}^2}\equiv \mu \partial b_{\varphi}^2/\partial\mu$, where $\mu$ is the mass scale
introduced in the renormalization procedure. Detailed expositions of the involved field theoretical calculations can be found in \cite{Ryder,Srednicki,Delamotte}.
By calculating the relevant correlation functions and
using the renormalization group equation, we obtain $\beta_{b_{\varphi}^2}=0$.
That is nothing else but the reflex of the Lorentz symmetry of the model. The speed of light is not an effective parameter in
the sense of the renormalization group. That is the reason we are free to choose the natural system of units or
any another one.
The renormalization group analysis is completely unnecessary. In fact, the Lagrangian
 (\ref{q1a}) just correspond to a units system where the speed of light is not equal one. Of course,
 this does not violate the Lorentz symmetry. The factor $b_{\varphi}$ can be eliminated by means of a rescaling
 of the spatial coordinates.

The situation is different when we have more than one single field in the theory since there is the possibility
of assigning different coefficients for the terms involving the spatial derivatives.
To be concrete, consider the Yukawa-type model in $3+1$ spacetime dimensions \cite{Gomes},
\begin{eqnarray}
\mathcal{L}&=&\frac12 \partial_{0}\varphi \partial_{0}\varphi-
\frac{b_{\varphi}^2}{2} \partial_{i}\varphi \partial_{i}\varphi-\frac{m^2}{2}\varphi^2
+\bar{\psi}(i\gamma^0\partial_0+i b_{\psi}\gamma^i\partial_i-M)\psi\nonumber\\
&+&ig\bar{\psi}\gamma^5\psi\varphi  -\frac{\lambda}{4!}\varphi^4,
\label{q2}
\end{eqnarray}
where $\psi$ and $\bar{\psi}\equiv\psi^{\dagger}\gamma^0$ are Dirac spinors and $\gamma^{\mu}$ are the Dirac matrices, with 
$\gamma^5\equiv i\gamma^0\gamma^1\gamma^2\gamma^3$. 
Contrarily to the previous case, it is no longer possible simultaneously to eliminate the parameters
$b_{\varphi}$ and $b_{\psi}$ by means of a rescaling of the spatial coordinates.
Hence, we have a true Lorentz symmetry breaking. In such situation the behavior of the effective
parameters is nontrivial. It is clear that when
$b_{\varphi}=b_{\psi}$, the renormalization group functions $\beta_{b_{\varphi}^2}$ and $\beta_{b_{\psi}}$
vanish since it corresponds to the case of the Lorentz invariant theory, i.e., the Lorentz symmetry manifests as a fixed point.
The important question here is concerning the nature of that fixed point.

\begin{figure}[!h]
\centering
\includegraphics[scale=0.8]{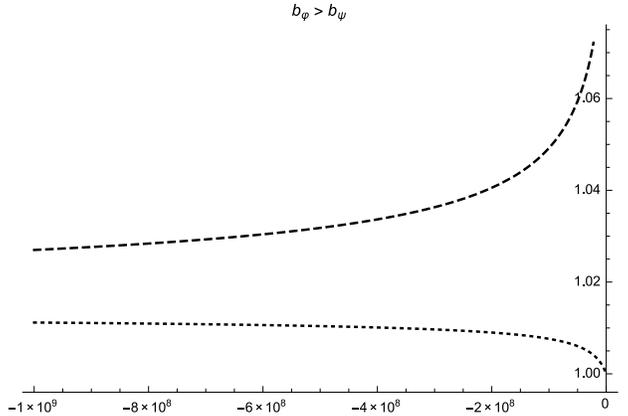}
\caption{Flux of parameters with the initial configuration $b_{\varphi}>b_{\psi}$. $b_{\varphi}(t)$ is given by  the dashed line
while $b_{\psi}(t)$ by the dotted line, with $t\equiv\ln \frac{\mu}{\mu_0}$ and $\mu_0$ ($t=0$) being the scale corresponding to
the initial configuration in the parameter space. The plot corresponds to the initial values $b_{\varphi}(0)=1.1$,
$b_{\psi}(0)=1$ and $g=0.001$.}
\label{rg1}
\end{figure}

By calculation the Green functions and using the renormalization group equation, we get
\begin{equation}
\beta_{b_{\varphi}^2}=\frac{1}{4\pi^2}\frac{(b_{\varphi}+b_{\psi})}{b_{\psi}^3}(b_{\varphi}-b_{\psi})g^2
\label{q3}
\end{equation}
and
\begin{equation}
\beta_{b_{\psi}}=-\frac{1}{6\pi^2}\frac{1}{b_{\varphi}(b_{\varphi}+b_{\psi})^2}(b_{\varphi}-b_{\psi})g^2.
\label{q4}
\end{equation}
As expected, when $b_{\varphi}=b_{\psi}$, we have $\beta_{b_{\varphi}^2}=\beta_{b_{\psi}}=0$.
Now let us analyze the situation where $b_{\varphi}\neq b_{\psi}$. There are two possibilities:

1. $b_{\varphi}>b_{\psi}$. In this case, the function ${b_{\varphi}}(t)$ is monotonically increasing while
${b_{\psi}}(t)$ monotonically decreasing, where $t$ is a logarithmic scale, $t\equiv\ln\frac{\mu}{\mu_0}$, with $\mu_0$ corresponding to
the initial configuration of the parameters.
As we go to lower and lower energies, the parameters tend to
a common value. This behavior is illustrated in Fig. \ref{rg1}.

\begin{figure}[!h]
\centering
\includegraphics[scale=0.8]{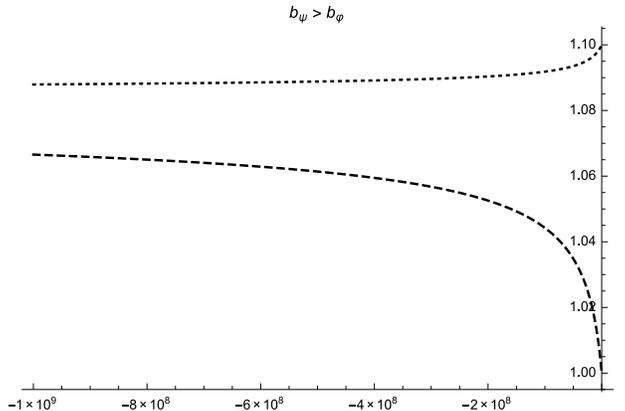}
\caption{Flux of parameter with the initial configuration $b_{\varphi}<b_{\psi}$. The plot corresponds to the initial values $b_{\varphi}(0)=1$,
$b_{\psi}(0)=1.1$ and $g=0.001$.}
\label{rg2}
\end{figure}

2. $b_{\varphi}<b_{\psi}$. Here the behavior of the effective parameters is reversed.
The function ${b_{\varphi}}(t)$ becomes monotonically decreasing while
${b_{\psi}}(t)$ becomes monotonically increasing. Again, as we go to lower and lower energies, they tend to a common value, as
showed in Fig. \ref{rg2}.

\begin{figure}[!h]
\centering
\includegraphics[scale=0.7]{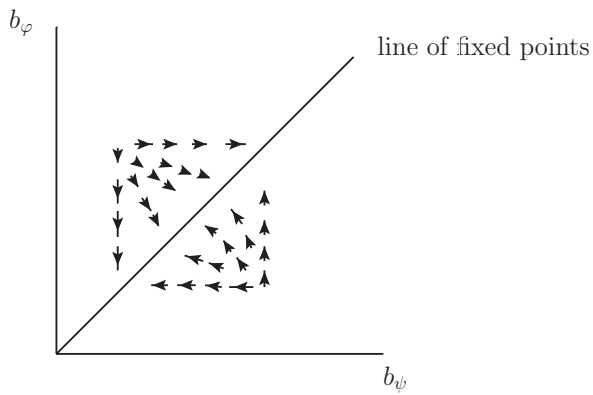}
\caption{Flow diagram of the paremeters $b_{\varphi}$ and $b_{\psi}$. There is a line of fixed points 
in $b_{\varphi}=b_{\psi}$.}
\label{rgflow}
\end{figure}

These results show that no matter what is the initial configuration of the parameters, as we go to the low-energy
sector of the theory we are approaching to the Lorentz invariant situation, showing that the fixed point is an attractive one. See the 
diagram of flow in Fig. \ref{rgflow}. In other words, the Lorentz symmetry is an inevitable consequence of the renormalization group
as we go to the low-energy limit.

%%%%%%%%%%%%%%%%%%%%%%%%%%%%%%%%%%%%%%%%%%

\section{Soft vs. Hard breaking of Symmetries}\label{SoftHard}

The modified Lagrangian (\ref{q2}), with two different parameters
$b_{\varphi}$ and $b_{\psi}$, is a close relative of the relativistic Lagrangian,
obtained when $b_{\varphi}=b_{\psi}$.
It corresponds to a kind of a soft deviation of the relativistic case. We say {\it soft} meaning that it just involves a
deformation of the parameters of the relativistic case but
the operator content of the theory is not changed.
However, once we no longer have the Lorentz symmetry,
we could think in introducing new symmetry-breaking marginal or relevant operators like, $b^{\mu}\varphi^2\partial_{\mu}\varphi$,
with $b_{\mu}$ a Lorentz-violating constant vectors.
This type of operator introduces a more drastic breaking of the Lorentz symmetry, since
it modifies the operator content of the theory. We call this a {\it hard} symmetry breaking.

For symmetries in general, we can formalize a bit more the definitions of
soft and hard symmetry breaking as follows.
Consider a Lagrangian of form (\ref{3.5}), that we rewrite here for convenience,
\begin{equation}
\mathcal{L}=\sum_i g_i\mathcal{O}_i.
\label{5.1}
\end{equation}
This Lagrangian is suppose to exhibit a certain set of symmetries, which means that there are
specific conditions or relations between the parameters $g_i$'s, that can be written as
\begin{equation}
F_m(g_i)=0, ~~~\text{with}~m\leq i_{max}.
\label{5.2}
\end{equation}
We obtain a Lagrangian with a soft breaking\footnote{This terminology is different from that one usually found
in the literature for the soft breaking of symmetry, which corresponds to breaking of symmetry by relevant operators, i.e., operators 
$\mathcal{O}$ with $[\mathcal{O}]<D$.}  of symmetry by replacing the parameters $g_i$ by
a set of deformed parameters, $\tilde{g}_i$, that do not satisfy all relations in (\ref{5.2}),
\begin{equation}
\mathcal{L}_{\it soft}=\sum_i \tilde{g}_i\mathcal{O}_i.
\label{5.3}
\end{equation}
The other possibility is to introduce in the Lagrangian new marginal and relevant operators which break
some symmetries
\begin{equation}
\mathcal{L}_{\it hard}=\sum_i g_i\mathcal{O}_i+\sum_j\lambda_j \tilde{\mathcal{O}}_j.
\label{5.4}
\end{equation}
This is defined as a hard breaking of the symmetries.
Finally, we can have the situation where both symmetry-breaking possibilities occur at the same time,
\begin{equation}
\mathcal{L}_{\it soft+hard}=\sum_i \tilde{g}_i\mathcal{O}_i+\sum_j\lambda_i \tilde{\mathcal{O}}_j,
\label{5.5}
\end{equation}
We say in this case that we have  {\it soft}+{\it hard} symmetry breaking.

After we give up some symmetry, the natural question is whether it can be reobtained or emerge in specific sectors of
the theory, for example in low-energies. With this type of mechanism in mind we can think in interesting extensions of several
physical models.
In the case of soft symmetry breaking, there are obvious fixed points, corresponding to the symmetric case with $\tilde{g}_i= g_i$,
as in the previous section.
One important question in this case concerns the nature of the fixed points, i.e., if they are attractive or repulsive.
On the other hand, in the case of hard or even soft+hard the situation is very delicate since it may involve
stringent fine-tuning. The existence of infrared fixed points is highly nontrivial in these cases.
To make the above general discussion a bit more concrete we consider some examples in the following.

%%%%%%%%%%%%%%%%%%%%%%%%%%%%%%%%%%%%%%%%%
\subsection{Hard Breaking of Lorentz Symmetry}\label{HardLorentz}

A simple example of a theory with a hard breaking of the Lorentz symmetry is provided by a deformation of (\ref{q1a}),
reducing the Lorentz symmetry down to the rotational invariance,
\begin{equation}
\mathcal{L}=\frac12 \partial_{0}\varphi \partial_{0}\varphi-\frac{b_{\varphi}^2}{2} \partial_{i}\varphi \partial_{i}\varphi
-\frac{\alpha^2_{\varphi}}{2}\Delta\varphi\Delta\varphi
-\frac{m^2}{2}\varphi^2
-\frac{\lambda}{3!}\varphi^3,
\label{5.6}
\end{equation}
with $\Delta\equiv \partial_i\partial_i$. It modifies the theory in an essential way compared to the relativistic case.
The operators $\partial_{0}\varphi \partial_{0}\varphi$ and $\Delta\varphi\Delta\varphi$ induce an
anisotropy between the space and time coordinates, which can be taken into account by assigning the dimensions, 
\begin{equation}
[x^0]=2~~~\text{and}~~~[x^i]=1.
\label{5.8}
\end{equation}
The dimension of the scalar field in this case is $[\varphi]=(d-2)/2$ and of the coupling constant is $[\lambda]=(10-d)/2$, 
with $d$ being the spatial dimension. It follows that  
the model is strictly renormalizable in 10+1 dimensions, where
the operator $\Delta\varphi\Delta\varphi$ is marginal while $\partial_{i}\varphi \partial_{i}\varphi$ is relevant.
Now the low-energy sector will eventually involve both $\Delta\varphi\Delta\varphi$ and $\partial_{i}\varphi \partial_{i}\varphi$
operators, such that finding a Lorentz invariant region is a delicate task.
It depends on the dominance of the operator $\partial_{i}\varphi \partial_{i}\varphi$ over $\Delta\varphi\Delta\varphi$.
At the same time we cannot simply ignore the higher spatial derivative operator since we end up with a nonrenormalizable
theory. On dimensional grounds we can expect this pattern, but a careful renormalization group analysis is necessary to get a more accurate
conclusion.

We want to discuss some reasons for constructing theories with hard breaking of the Lorentz symmetry as in (\ref{5.6}).
The fundamental point is that the presence of higher spatial derivatives operators yield to an improvement of the ultraviolet behavior.
Hence the model turns out to be renormalizable even in 10+1 dimensions.
Of course, we could try to obtain the same gain without losing the Lorentz symmetry
by introducing the relativistic operator $\partial^2\varphi \partial^2\varphi$, with $\partial^2\equiv \partial_{\mu}\partial^{\mu}$.
The problem here is that this operator generally leads to a breakdown of the unitarity, which is a key property of a
consistent quantum theory. This problem is directly related to the presence of higher time derivatives which essentially modifies the pole structure of the
Green functions leading ultimately to the violation of the cutting rules (Cutkosky) required for unitarity \cite{Cutkosky}.
Thus (\ref{5.6}) is a natural way to get a theory with a better ultraviolet behavior without spoiling unitarity at the price of the
Lorentz symmetry, which is no longer a property of the full spectrum but could be restricted to an specific range of energy.
After all, within the current experimental limitations, the constraints on Lorentz-violating parameters are very strong \cite{Russell}.

With that understanding it is tempting to employ a similar procedure to the problem of quantization of gravitational
interactions in the perturbative sense. We will discuss more about gravitation in Sec. \ref{Discrete}, but
for now the important point is that the Einstein-Hilbert action for the gravitational field is not perturbatively renormalizable in 3+1 spacetime dimensions.
With the above procedure, we could obtain a power-counting renormalizable anisotropic quantum gravitational theory in 3+1 dimensions \cite{Horava}.
There are not yet clear conclusions if this is a viable mechanism but it could at least furnish more insights on this formidable problem.

To conclude this section, it is instructive to use these ideas to exemplify the soft+hard breaking of Lorentz
symmetry in the Yukawa model of Sec. \ref{Lorentz}.  This can be done just by adding the operators
$\Delta\varphi\Delta\varphi$ as well as $\bar{\psi}(i\gamma^i\partial_i)^2 \psi= \bar{\psi}\Delta \psi$ to the Lagrangian (\ref{q2}).
This induces the anisotropic scaling between space and time which shifts the renormalizability to 6+1 spacetime dimensions.
Renormalization group studies of this model indicate that it is hard to clearly identify
a Lorentz-invariant sector, although it can be reached after stringent fine-tuning of the involved parameters \cite{Gomes}.

%%%%%%%%%%%%%%%%%%%%%%%%%%%%%%%%%%%%%%%%%%%
\subsection{Soft Breaking of Supersymmetry}\label{SoftSUSY}

The notion of soft breaking of symmetry can be used to investigate different types of symmetry.
Let us apply it to the case of supersymmetry by considering
a soft-deformed version of the massless Wess-Zumino model discussed in \cite{Iliopoulos},
\begin{equation}
\mathcal{L}=\frac{1}{2}\partial_{\mu}A\partial^{\mu}A+\frac{1}{2}\partial_{\mu}B\partial^{\mu}B+
\frac{1}{2}\bar{\psi}i\gamma^{\mu}\partial_{\mu}\psi-f(\bar{\psi}\psi A+i\bar{\psi}\gamma_5 \psi B)
-\frac{1}{2} \lambda (A^2-B^2)^2-2gA^2B^2.
\label{5.9}
\end{equation}
This Lagrangian is invariant under supersymmetry transformations when $\lambda=g=f^2$. We will analyze the renormalization group flow
of the couplings constants $\lambda$, $g$ and $f$. The new feature of the theory is that the Ward
identities resulting from supersymmetry are no longer valid. This implies, for example, 
that the well known canceling of logarithmic divergences of the interaction vertices   
and the quadratic divergences in the massive case will no longer take place.
The corresponding $\beta$-functions are given by
\begin{equation}
\pi^2\beta_{f^2}=\frac{3}{2} f^4,
\label{5.10}
\end{equation}
\begin{equation}
\pi^2\beta_{\lambda}=f^2\lambda +\frac{9}{4}\lambda^2+\frac{1}{4}g^2-2 f^4
\label{5.11}
\end{equation}
and
\begin{equation}
\pi^2\beta_{g}=f^2\lambda +\frac{3}{2}\lambda g+g^2-2 f^4.
\label{5.12}
\end{equation}
To make the nature of the fixed point more evident, it is
convenient to introduce two parameters to measure the deviations from supersymmetry,
\begin{equation}
\delta_1\equiv \frac{\lambda-g}{f^2}~~~\text{and}~~~\delta_2\equiv \left(\frac{\lambda}{f^2}-1\right)+
\frac{1}{5}\left(\frac{g}{f^2}-1\right).
\label{5.13}
\end{equation}
The linearized $\beta$-functions for $\delta_1$ and $\delta_2$ are
\begin{equation}
\beta_{\delta_1}\sim \frac{3}{2}\delta_1 ~~~\text{and}~~~\beta_{\delta_2}\sim\frac{9}{2}\delta_2.
\label{5.14}
\end{equation}
The positive signs show that the supersymmetric point $\delta_1=\delta_2=0$ is attractive in the infrared.
Thus the supersymmetry is a low-energy manifestation of the deformed model (\ref{5.9}).

%%%%%%%%%%%%%%%%%%%%%%%%%%%%%%%%%%%%%%%%%%%
\subsection{Soft+Hard Breaking of Gauge Symmetry}\label{Gauge}

Another interesting application involves gauge invariance.
We consider the scalar electrodynamics with different coupling constants yielding to a soft deformation from the
gauge-invariant case,
\begin{equation}
\mathcal{L}=\partial_{\mu}\varphi\partial^{\mu}\varphi^*-m^2\varphi \varphi^*-\frac{1}{4}F_{\mu\nu}F^{\mu\nu}
+ie A^{\mu}(\varphi^*\partial_{\mu}\varphi-\varphi\partial_{\mu}\varphi^* )+\alpha\varphi\varphi^*A_{\mu}A^{\mu}
-\frac{\lambda}{4}(\varphi^*\varphi)^2.
\label{5.15}
\end{equation}
As the breaking of the gauge invariance occurs in the interaction part, we include the standard gauge-fixing term,
\begin{equation}
\mathcal{L}_{g.f.}=-\frac{\xi}{2}(\partial_{\mu}A^{\mu})^2
\label{5.16}
\end{equation}
and work in the Feynman gauge, $\xi=1$.
When $e^2=\alpha$ the Lagrangian (\ref{5.15}) is gauge-invariant. Without the protection of the gauge invariance, operators like
\begin{equation}
\mathcal{L}_{hard}=-\frac{m^2}{2}A_{\mu}A^{\mu}+\frac{g}{4}(A_{\mu}A^{\mu})^2,
\label{5.17}
\end{equation}
are expected to ensure renormalizability. Thus the soft breaking induces a hard breaking of the gauge invariance.
It is convenient to work with the variables
\begin{equation}
\delta_1\equiv \frac{\alpha}{e^2}-1, ~~~\delta_2\equiv \frac{g}{e^2}, ~~~\text{and}~~~\Delta\equiv \frac{\lambda}{e^2}.
\label{5.18}
\end{equation}
Up to linear order in both $\delta_1$ and $\delta_2$ the $\beta$-functions are \cite{Iliopoulos},
\begin{equation}
\beta_{e^2}\sim \frac{1}{3}e^4,~~~\beta_{\delta_1}\sim e^2[(\Delta-3)\delta_1-\frac{9}{2}\delta_2],~~~
\beta_{\delta_2}\sim \frac{1}{3}e^2\delta_2,~~~\beta_{\Delta}\sim e^2[12+\frac{5}{2}\Delta^2-\frac{19}{3}\Delta].
\label{5.19}
\end{equation}
Note that $\beta_{\Delta}>0$. Thus in the low-energies we eventually get $\Delta\ll 3$. The positive sign of $\beta_{\delta_2}$
tells that $\delta_2\rightarrow 0$ in that limit. In this situation $\beta_{\delta_1}$ tends to
\begin{equation}
\beta_{\delta_1}\sim -e^2[3\delta_1+\frac{9}{2}\delta_2],
\label{5.20}
\end{equation}
showing that the infrared fixed point is repulsive, i.e., the gauge invariance is not a low-energy property of the system.

%%%%%%%%%%%%%%%%%%%%%%%%%%%%%%%%%%%%%%%%

%%%%%%%%%%%%%%%%%%%%%%%%%%%%%%%%%%%%%%%%%%

\section{Scale Invariance at Critical Points}\label{CP}

One of the most important and interesting example of an emergent symmetry is the scale invariance
when a system undergoes a second order phase transition. As we shall discuss, this symmetry is responsible for the most
striking  properties of the critical points, such as the universality. By exploring the consequences of the scale invariance
it is possible to extract a lot of useful information.

Different phases of a system can be characterized according to their symmetries (or absence of them), giving rise to  
the idea of order classification. 
When the system undergoes a phase transition it changes its physical properties, i.e., it changes the order.
To characterize this process, it is usual to introduce a quantity, called order parameter, with the following
properties: in one of the phases it vanishes ({\it disordered} phase), while it is nonzero in the other phase ({\it ordered} phase). Furthermore, it
generally reflects the symmetry properties of the system.
We are restricting our discussion to the case of continuous (second order) phase transitions. Near the critical point, the order parameter is
small, justifying an expansion of the free energy in powers of the order parameter, known as
Landau-Ginzburg free energy (see for example \cite{Binney}). Of course, we are not going into the merit of how easy or hard it is to
find an order parameter.

The Landau-Ginzburg free energy provides a general description of the critical behavior for classes of systems. 
On the other hand, the phase transitions occur in many different
systems, with many different properties, each one with its own microscopic features.
An immediate question is: what is the reason for considering a general theory for phase transitions?
The answer leads to the concept of the universality, that is the most special  property
of the critical phenomena, and it is ultimately related to the emergence of the scale invariance at the
critical point.

In the neighborhood of the critical region some thermodynamics quantities, as the
specific heat and the susceptibility, exhibit a peculiar behavior, with divergences that are
governed by a set of critical exponents. In this sense, the critical exponents describe the nature of the singularities and
characterize the phase transition. All phase transitions with the same set of critical exponents belong to
the same universality class. The most remarkable property is that the universality classes are determined by few factors,
essentially the spatial dimension of the system and the involved symmetries.

What is behind the universality is the fact that the correlation length is the only
scale length of the theory at the critical point. Away from the critical point, the correlation length is of the order of the typical 
microscopic length of the system, as the lattice spacing in the case of a solid. Thus the microscopic properties are important and 
give rise to the specific characteristics of
the material. However, in the neighborhood of the critical region, the correlation length becomes larger and larger, 
diverging at the critical point (when it has the size of the system) such that
the microscopic properties become unimportant and the system scale invariant. This emergent symmetry can
be exploited to restrict the correlation functions.

Close to the critical point, the correlation length $\xi$ behaves as
\begin{equation}
\xi\sim |t|^{-\nu},
\label{6.1}
\end{equation}
where $\nu$ is the critical exponent of the correlation length and $t$ involves some parameter which
measures the distance from the critical point. For example, in a phase transition driven by the temperature,
it is usually choosen to be $t\equiv (T-T_c)/T_c$, with $T_c$ being the critical temperature.
Based in the scale invariance, the two-point correlation function of the order parameter is
\begin{equation}
G(r)\sim \frac{e^{-r/\xi}}{r^{d-2+\eta}},
\label{6.2}
\end{equation}
with $\eta$ being the critical exponent of the correlation function. Note that away from the critical point,
the correlation function exhibits an exponential decay, while it becomes a power law at the critical point
($\xi\rightarrow\infty$), as required by the scale invariance.

According to Sec.\ref{SoftHard}, the mechanism
of emergence of scale invariance generally occurs when the parameters associated to the
hard symmetry-breaking operators, $\lambda_i$, vanish at the critical point. For example,
in the case of temperature-driven phase transitions, $\lambda_i\sim (T-T_c)$. These operators are generally associated
to the quadratic term in the order parameter of the Landau-Ginzburg expansion.

%%%%%%%%%%%%%%%%%%%%%%%%%%%%%%%%%%%%%%%%%%%%%%
\section{Ising Model and Conformal Invariance}\label{CI}

One of the most important models of statistical mechanics is the Ising model, described by the Hamiltonian,
\begin{equation}
\mathcal{H}=-J\sum_{<i,j>}\sigma_i\sigma_j-h\sum_i \sigma_i,
\label{7.1}
\end{equation}
where $\sigma_i=\pm1$ and $h$ is an external field. As firstly obtained by Onsager \cite{Onsager}, 
the two dimensional case is exactly solvable in the absence of the external field and it exhibits
an order-disorder continuous phase transition.
That is a perfect example to test the methods of effective field theory as we can always compare its implications 
with those extracted from the exact solution. The basic question is: how can we obtain an effective field
theory for the Ising model at the critical point, where the continuum limit is legitimate?
The answer is conformal invariance. In fact, as it can be shown \cite{Polchinski2,Nakayama},
under certain broad conditions, the scale invariance in two dimensions implies conformal invariance.
There is no general proof of this statement in higher dimensions although there are no counterexamples. Anyway,
conformal invariance does not bring much more information than rigid scale invariance in three  or higher dimensions\footnote{Recently, there has been much interest on the constraints of conformal invariance in three dimensions \cite{Poland}.}.
This is not the case of two dimensional systems, where the conformal invariance is strong enough, leading
to severe restrictions on the operator content as well as on the correlation functions of the theory.
We will explore the consequences of conformal symmetry.

Let us start by stressing some aspects of conformal symmetry important to our purposes.
For a detailed exposition see the standard references \cite{Ginsparg,Francesco,Mussardo}.
A conformal transformation "preserves angles", i.e.,
it is a coordinate transformation that leaves the metric invariant up to a scale factor,
\begin{equation}
g_{\mu\nu}(x)\rightarrow g^{\prime}_{\mu\nu}(x^{\prime})=\Omega(x)g_{\mu\nu}.
\label{7.2}
\end{equation}
For an infinitesimal transformation, $x^{\mu}\rightarrow x^{\prime\mu}= x^{\mu}+\epsilon^{\mu}(x)$, under which the
interval $ds^2$ changes as
\begin{equation}
ds^2\rightarrow ds^2 + (\partial_{\mu} \epsilon_{\nu}+\partial_{\nu} \epsilon_{\mu})dx^{\mu}dx^{\nu},
\label{7.3}
\end{equation}
the requirement that (\ref{7.2}) is true leads to the following condition for the parameter $\epsilon^{\mu}$,
\begin{equation}
\partial_{\mu} \epsilon_{\nu}+\partial_{\nu} \epsilon_{\mu}=\frac{2}{D} (\partial\cdot\epsilon)\eta_{\mu\nu},
\label{7.4}
\end{equation}
where $\eta_{\mu\nu}$ is the flat spacetime metric, with Euclidean or Minkowski signature. We will work in Euclidean space. 
The solutions of this condition for $D>2$ give the transformations of the conformal group:
\begin{eqnarray}
&&x^{\mu}\rightarrow x^{\prime\mu}= x^{\mu}+a^{\mu} \Rightarrow \text{translations}\nonumber\\
&&x^{\mu}\rightarrow x^{\prime\mu}=\Lambda^{\mu}\,_{\nu}x^{\nu}\Rightarrow \text{Lorentz transformations}\nonumber\\
&& x^{\mu}\rightarrow x^{\prime\mu}=\lambda x^{\mu}\Rightarrow \text{dilatations}\nonumber\\
&&x^{\mu}\rightarrow x^{\prime\mu}=\frac{x^{\mu}+b^{\mu} x^2}{1+2 b\cdot x+b^2 x^2}\Rightarrow\text{special conformal transformations}.
\label{7.5}
\end{eqnarray}
All the above parameters are real. 
In the case of translations and special conformal transformations they are constant vectors, $a^{\mu}$ and $b^{\mu}$, respectively. Lorentz 
transformations involve a matrix $\Lambda^{\mu}\,_{\nu}$ and dilatations corresponds to a multiplication by a scalar constant
parameter $\lambda$.
Notice that in two dimensions the condition (\ref{7.4})  becomes
\begin{equation}
\partial_1\epsilon_1=\partial_2\epsilon_2~~~\text{and}~~~\partial_1\epsilon_2=-\partial_2\epsilon_1,
\label{7.6}
\end{equation}
that are nothing else but the Cauchy-Riemann conditions for an analytic complex function.
That means that the conformal transformations are equivalent to the transformations of analytic functions
\begin{equation}
z\rightarrow f(z)~~~\text{and}~~~\bar{z}\rightarrow \bar{f}(\bar{z}),
\label{7.7}
\end{equation}
with $z=x^1+i x ^2$ and $\bar{z}=x^1-i x ^2$.
To generate arbitrary analytic transformations we need an infinite set of generators, i.e., the
conformal group in two dimensions has infinite generators. This is not so in higher dimensions.

Now we have to discuss some consequences of the two-dimensional conformal invariance in the quantum theory.
As we discussed in Sec. \ref{EFT}, a general property of quantized fields is the singular nature of the product of operators at very close spacetime points.
In general, given two quantized operators $\Phi_1(x)$ and $\Phi_2(y)$,
the singularities are encoded into an operator product expansion (OPE),
\begin{equation}
\Phi_1(x) \Phi_2(y)\sim \sum_i f_i(x-y) \mathcal{O}_i(y),
\label{7.7a}
\end{equation}
where the coefficients $f_i(x-y)$ are singular when $x\rightarrow y$ and $\mathcal{O}_i$ is a set o local operators.
On dimensional grounds, we see that
\begin{equation}
f_i(x-y)\sim \frac{1}{|x-y|^{[\Phi_1]+[\Phi_1]-[\mathcal{O}_i]}}.
\label{7.7b}
\end{equation}
The study of OPE's properties are of primary interest since they allow us to obtain several correlation functions of
the theory, given by the expected values of the product of operators.
All relevant physical information are ultimately contained into the correlation functions.

We are interested in a particular class of operators, called primary operators,
that under (\ref{7.7}) transforms as
\begin{equation}
\Phi(z,\bar{z})\rightarrow \left(\frac{\partial f}{\partial z}\right)^h \left(\frac{\partial \bar{f}}{\partial \bar{z}}\right)^{\bar{h}}
\Phi(f(z),\bar{f}(\bar{z})).
\label{7.8}
\end{equation}
The parameters $(h,\bar{h})$ are called the conformal weights of the primary operator $\Phi(z,\bar{z})$.
This transformation property can be put in terms of an OPE algebra of the form (\ref{7.7a}).
The idea is to identify the generator of conformal transformations. Given a conserved traceless energy-momentum tensor, $T_{\mu\nu}$,
we can immediately construct the conformal current as $j^{\mu}=T^{\mu\nu}\epsilon_{\nu}$,
with $\epsilon_{\mu}$ satisfying (\ref{7.4}). The transformation is generated by the
conserved charge $Q\equiv \int d^dx j^0$.
This shows that the energy-momentum is essentially the generator of conformal transformations.
By writing the components of the energy momentum in complex coordinates, $T(z)$ and $\bar{T}(\bar{z})$, it can be shown that
the OPE of a primary field with $T(z)$ is \cite{Ginsparg}
\begin{equation}
T(z)\Phi(w,\bar{w})=\frac{h}{(z-w)^2}\Phi(w,\bar{w})+\frac{1}{z-w}\partial_w \Phi(w,\bar{w})+\cdots
\label{7.9}
\end{equation}
and a similar expression for the OPE with $\bar{T}(\bar{z})$ replacing $h\rightarrow\bar{h}$.
This is an equivalent way to define a primary operator.
In other words, a primary operator of conformal weight $(h,\bar{h})$ has the above OPE with the energy-momentum components.
Two important quantities associated to an operator are the scaling dimension, $\Delta=h+\bar{h}$, and
the spin, $s=h-\bar{h}$. This is so since they are the eigenvalues of the generators of scale and rotation transformations, respectively.
For example, scalar operators are given by $h=\bar{h}$. These quantities will be important later, when we will identify the
operator content of the statistical models.

Not all operators transform according (\ref{7.8}) and, consequently, they will not satisfy (\ref{7.9}).
An important example is the own energy-momentum tensor. For the $T(z)$ component, for example, we have,
\begin{equation}
T(z)T(w)=\frac{c/2}{(z-w)^4}+\frac{h}{(z-w)^2}T(w)+\frac{1}{z-w}\partial_w T(w)+\cdots.
\label{7.10}
\end{equation}
The constant parameter $c$ is the {\it central charge} and plays a fundamental role in conformal field theories.
It can be interpreted as a quantum anomaly, in the sense that it implies a violation of the primary field condition (\ref{7.9})
whenever $c\neq 0$. Since $\langle T(z)T(0)\rangle=(c/2)/z^4$,
we expect that unitary theories have $c\geq 0$.

It is instructive to discuss the realization of the conformal symmetry as fixed points of the renomalization group.
This is the content of the Zamolodichkov's $c$-theorem \cite{Zamolodchikov}.
The basic idea underlying the $c$-theorem is that the procedure of integrating out over
high-energy degrees of freedom, as discussed in the Sec. \ref{EFT}, is irreversible since we are losing information.
It must exist some function which is monotonic decreasing along the renormalization group flow.
Thus the $c$-theorem establishes the existence of a function $C$, depending on the coupling constants of the theory, with the property
\begin{equation}
\frac{dC}{dt}\leq 0,
\label{7.11}
\end{equation}
along the renormalization group flow, with $t=\log (\mu |z|^2)$ and $\mu$ being a mass scale.
It is stationary at the fixed points and its value coincides with the central charge, $C^{\ast}=c$.

As an aside, note that the physical reasoning behind the $c$-theorem is independent of spacetime dimension, but the
two-dimensional proof cannot be extended to higher dimensions. Only recently 
a four-dimensional proof of the $c$-theorem was obtained \cite{Komargodski}.

There is a particular class of unitary theories, with the central charge restricted to the interval $0< c < 1$,
that possesses a finite set of primary operators. They are called {\it minimal models} and, for consistence,
the central charges are enforced to assume discrete values,
\begin{equation}
c(m)=1-\frac{6}{m(m+1)},~~~\text{with}~m=3,4,5,\cdots,
\label{7.12}
\end{equation}
while the conformal weights are constrained to be
\begin{equation}
h_{p,q}(m)=\frac{[(m+1)p-mq]^2-1}{4m(m+1)},
\label{7.13}
\end{equation}
with $p,q$ being integers satisfying $1\leq p\leq m-1$ and $1\leq q \leq m$.
Interesting enough, the first members of the series $m=3,4,5$ can be identified with some important statistical models, as the
Ising, Tricritical-Ising, and the 3-states Potts models, respectively.

Let us concentrate on the first member $m=3$. For it the central charge is
$c=1/2$ and the conformal weights are shown in the table \ref{table1}.
\begin{table}[htb]
\centering
\large
\begin{tabular}{|c|c|}
\hline
$1/2$ & 0  \\ \hline
$1/16$ & $1/16$    \\ \hline
$0$ &  $1/2$  \\ \hline
\end{tabular}
\caption{The entries of the table represent the values of $h$ for the allowed $q$'s and $p$'s for $m=3$, such that the
values of $q$ run in the vertical whereas the values of $p$ run in the horizontal.}
\label{table1}
\end{table}
The operator content exactly matches that one of the Ising model.
In particular, consider the scalar operators ($h=\bar{h}$) of the  form $\Phi_{p,q}(z,\bar{z})\equiv \phi_{p,q}(z)\bar{\phi}_{p,q}(\bar{z})$.
The conformal invariance constraints their correlation functions to behave as
\begin{equation}
\langle \Phi_{p,q}(z_1,\bar{z}_1)\Phi_{p,q}(z_2,\bar{z}_2)\rangle\sim \frac{1}{|z_1-z_2|^{2\Delta_{p,q}}},
\label{7.14}
\end{equation}
remembering that $\Delta_{p,q}=h_{p,q}+\bar{h}_{p,q}$.
We have three operators with the respective conformal weights,
\begin{equation}
\Phi_{1,1}: (0,0),~~~\Phi_{2,1}: (1/2,1/2),~~~ \Phi_{1,2}: (1/16,1/16).
\label{7.15}
\end{equation}
$\Phi_{1,2}$ is identified as the local magnetization $\sigma(x)$ while $\Phi_{2,1}$ is identified
as the local energy density $\epsilon(x)$. At the critical point, the two-point correlation function for the magnetization is given by (\ref{6.2}),
\begin{equation}
\langle \sigma (x_1)\sigma (x_2)\rangle \sim \frac{1}{|x_1-x_2|^{  d-2+\eta}},
\label{7.16}
\end{equation}
while for the local energy density it can be expressed in terms of the exponent $\nu$ introduced in (\ref{6.1}),
\begin{equation}
\langle \epsilon (x_1)\epsilon (x_2)\rangle \sim \frac{1}{|x_1-x_2|^{2(d-\frac{1}{\nu})}}.
\label{7.17}
\end{equation}
By comparing with (\ref{7.14}) we obtain the critical exponents $\eta =1/4$ and $\nu=1$,
that are exactly the critical exponent obtained from exact Onsager's solution.
This member has also fermionic operators with conformal weights (1/2,0) and (0,1/2).
Thus we can write down a fixed point action with central charge $c=1/2$ in terms of these operators
\begin{equation}
S^{\ast}=\int d^2z (\psi \bar{\partial}_{\bar z} \psi+\bar{\psi}\partial_z\bar{\psi} ),
\label{7.18}
\end{equation}
that are Majorana fermions. Thus we conclude that the effective field theory for the Ising model at the critical point can
be written in terms of a fermionic fields.

Before closing this section let us summarize what we have done.
We started by looking for an effective field theory for the Ising model at the critical point. We then invoke the scale invariance.
It turns out that the scale invariance in two dimensions implies conformal invariance. The remarkable feature is
that the two-dimensional conformal symmetry is so powerful that give us much more, i.e.,
instead of obtain only the effective field theory for the Ising model, we obtain a class of theories (minimal models) that
can be identified with several statistical models.
Of course, in general the symmetries are not so powerful as conformal symmetry
but yet they constitute maybe the best guide to furnish physical information.

%%%%%%%%%%%%%%%%%%%%%%%%%%%%%%%%%%%%%%%%%%%%%%%%%%

\section{Accidental Symmetries}\label{AS}

Some emergent symmetries do not follow from any physical property of the system,
but just from an accidental (and of course special) combination of the relevant degrees of freedom acting in determined scale.
This will be translated to the resulting effective field theory in a form of an {\it accidental} symmetry.

To exemplify we can imagine the following situation. Suppose that we have a system
involving both bosonic and fermionic constituents. In fact, there are innumerable physical systems that are
composed by both bosons and fermions, or some degrees of freedom that effectively behave as bosons and fermions.
In general such systems depend on a set of controllable parameters such that
it could be that for certain values of these parameters, or in certain region of the parameter space,
they turn out to have the same status, for example, the same effective masses.
If this is the case, the model will exhibit an accidental supersymmetry.

We want to discuss some examples where this type of mechanism takes place, not only in the case of supersymmetry but
also involving Lorentz symmetry.

%%%%%%%%%%%%%%%%%%%%%%%%%%%%%%%%%%%%%%%%%%%%%
\subsection{Tricritical Ising Model and SUSY}

The conformal symmetry is not the only symmetry emerging at the critical point of the two-dimensional systems.
Surprisingly, the second member ($m=4$) of the minimal models  (\ref{7.12}) shows up a supersymmetric behavior.
In this case the conformal symmetry is enhanced to the superconformal symmetry.
We do not want to deepen in the superconformal invariance, but just to give a sketch on how this come up.
A nice reference on this subject is \cite{Qiu}.

The interesting point is that the second member $m=4$ can be identified as the tricritical Ising model at its tricritical point.
The tricritical point is the meeting point of a line of first order phase transition and of a line of second order phase transition.
The tricritical Ising model, also known as the Blume-Emery-Griffiths model, is a generalization of (\ref{7.1}) given by
\begin{equation}
\mathcal{H}=-J\sum_{<i,j>}t_it_j\sigma_i\sigma_j+\Delta\sum_it_i-h\sum_i \sigma_i,
\label{8.1}
\end{equation}
where $t_i=0,1$ represents the possibility of vacancies in the sites. It has been used in the study of magnetic systems as well as
in describing $^3\text{He}$-$^4\text{He}$ mixtures \cite{Blume}.
The odd property of the model is that it
exhibits a tricritical point for specific values of the parameters $(J_c,\Delta_c,h=0)$.

The intuitive idea on how the supersymmetry arises in this problem goes as follows. We saw that the conformal field theory for the Ising model is
given in terms of Majorana fermions (\ref{7.18}). In the tricritical Ising model, in addition to the $\sigma$-variables
we have the $t$-variables, and then we expected to have a more general action than (\ref{7.18}).
Heuristically, we could think that the $\sigma_i$'s are responsible for fermionic-type contributions to the action while
$t_i$ are responsible for bosonic-type contributions. At the tricritical point, the bosonic and fermionic contributions
acquire the same status such that the theory turns out to be supersymmetric.
\begin{table}[htb]
\centering
\large
\begin{tabular}{|c|c|c|}
\hline
$3/2$ & 7/16 & 0  \\ \hline
$3/5$ & $3/80$& 1/10    \\ \hline
$1/10$ &  $3/80$&  3/5 \\ \hline
$0$ & 7/16 & 3/2 \\ \hline
\end{tabular}
\caption{Conformal weights for the model $m=4$. }
\label{table2}
\end{table}

At a more technical level, the second member of the minimal model series (\ref{7.12}) is equivalent to the
first member of the minimal unitary superconformal serie. The central charge of the theory is $c=7/10$
and, according to (\ref{7.13}),  the allowed conformal weights for $m=4$ are shown in the table \ref{table2}.
Some of the primary operators possessing same properties under $Z_2$ transformation ($Z_2$ even fields)
can be combined into a superfield of conformal weight $(1/10,1/10)$. In fact, the
energy operator $\epsilon(z,\bar{z})$, with conformal weight $(1/10,1/10)$, the
vacancy operator $t(z,\bar{z})$, with conformal weight $(3/5,3/5)$,
together with the fermionic operators $\psi$ and $\bar{\psi}$ of conformal weight $(3/5,1/10)$,
can be seen as component fields of the superfield
\begin{equation}
\mathcal{N}(z,\bar{z},\theta,\bar{\theta})=\epsilon(z,\bar{z})+\bar{\theta}\psi(z,\bar{z})+\theta\bar{\psi}(z,\bar{z})+i\theta\bar{\theta} t(z,\bar{z}),
\label{8.2}
\end{equation}
where $\theta$ and $\bar{\theta}$ are the Grassmanian superspace coordinates.
A superspace action for $\mathcal{N}$ is
\begin{equation}
S=\int d^2z d^2\theta \left( \frac{1}{2}D\mathcal{N} \bar{D}\mathcal{N}+\mathcal{N}^3 \right),
\label{8.3}
\end{equation}
with the supercovariant derivatives defined as
\begin{equation}
D\equiv\frac{\partial}{\partial\theta}-\theta\frac{\partial}{ \partial z}~~~\text{and}~~~
\bar{D}\equiv\frac{\partial}{\partial\bar{\theta}}-\bar{\theta}\frac{\partial}{ \partial \bar{z}}.
\label{8.4}
\end{equation}
This example illustrates very well how an accidental combination of degrees of freedom
can be reflected as a symmetry in the effective field theory.
The important conclusion is that we can take advantage of the accidental symmetry to explore some properties of the system.
We close this discussion by noting that the tricritical Ising model constitutes the first example of a supersymmetric
field theory in nature, since it can be experimentally realized in the
adsorption of helium on krypton-plated graphite \cite{Ferreira}.

%%%%%%%%%%%%%%%%%%%%%%%%%%%%%%%%%
\subsection{SUSY at the Boundary of a Topological Phase}

The system in question is the (1+1)-dimensional edge of a time-reversal invariant (2+1)-dimensional topological superconductor
presented in \cite{Grover}.
We will go back to say some words on topological phases of matter in the end of the next section.
Now it is enough for our purposes to know that these topological phases typically support gapless modes at their boundaries.
For topological superconductors they are Majorana fermions which are protected by
the time-reversal symmetry from acquiring mass.
Spontaneous symmetry-breaking of this symmetry is a natural mechanism to gap them out.

To construct a model representing that physics, we need two basic degrees of freedom: the obvious Majorana fermions
as well as a symmetry-breaking field to generate mass for the Majorana fermions.  We will consider the edge as
a discretized one-dimensional lattice. It is convenient since we can use the fact
that the two-dimensional Ising model (\ref{7.1}) can be mapped into a one-dimensional quantum mechanical problem
with the Hamiltonian \cite{Susskind},
\begin{equation}
H=-\sum_i \left(\mu_i^x+\lambda \mu_{i+1}^z \mu_i^z\right),
\label{9.1}
\end{equation}
where $\mu^x$ and $\mu^z$ are Pauli matrices.
It exhibits a phase transition when $\langle \mu^z\rangle $ acquires a nonzero expectation value.
Next we need to couple it to the Majorana fermions $\chi_j$.
We will consider that the Majorana fermions live at the sites $j$ while the Ising spins live at the links $j\pm \frac{1}{2}$.
The Hamiltonian is written as
\begin{equation}
H=-i\sum_j\left(1-g\mu^z_{j+\frac{1}{2}}\right)\chi_j\chi_{j+1}+\sum_j
\left[J\mu^z_{j-\frac{1}{2}}\mu^z_{j+\frac{1}{2}}-h\mu^x_{j+\frac{1}{2}}\right].
\label{9.2}
\end{equation}
We can identify a point where the system behaves supersymmetrically by analyzing the central charge as a function of $h$ with
both $J$ and $g$ fixed. According to the results from computational simulation of \cite{Grover}, they get the following pattern.
When $h$ is large, we are in a gapless phase, with the Majorana fermions propagating at the edge with central charge $c=1/2$.
In the gaped phase, for small $h$, there are no conformal degrees of freedom and so $c=0$.
There is an intermediate region where something interesting it happens. For a specific value of $h=h_c$
the central charge becomes $c=7/10$. We remember from previous section that this is exactly the central charge
of the tricritical Ising model at its tricritical point, which in turn is supersymmetric. So the effective action at this point
is well described by (\ref{8.3}) showing that supersymmetry is emergent at $h=h_c$.

%%%%%%%%%%%%%%%%%%%%%%%%%%%%%%%%%%%%
\subsection{Lorentz Symmetry in the Quantum Hall Effect}

The quantum Hall effect exhibits extraordinary physics, representing a landmark of modern condensed matter
physics \cite{Wen0,Wen1,Wen2}. It is essentially constituted of electrons moving
on a two-dimensional surface under the presence of a strong perpendicular magnetic field.
While the physical setup is simple, the resulting phenomena are rather surprising.
They arise from a combination of different physical mechanisms. The magnetic field enforces the electrons to describe
circular orbits. Along their motion, eventually they go around at each other producing quantum phases.
Furthermore, the electrons tend to stay away as much as possible from each other due to strong Coulomb repulsion and
compliance with the Pauli's exclusion principle.
These ingredients together generate some highly coordinate patterns responsible for
very rich physical properties. The striking imprint is the plateau structure of the resistivity $\rho_{xy}\sim \frac{1}{\nu}$, 
with $\nu= 1,2,3,...,1/3,1/5,2/5,...$, 
where $\nu$ is the filling fraction given by the number of electrons per the number of states in each level of 
energy (Landau levels). 
This is typically a strongly correlated system that cannot be approached with the
traditional perturbative methods. There are different ways to probe the system but we will
focus on the effective field theory approach.
In addition to the appearance of Lorentz invariance as an accidental symmetry, this example
illustrates the power of the effective field theory methods of Sec. \ref{EFT} in obtaining low-energy
descriptions. A nice discussion on this is presented in \cite{Zee}.

To construct the low-energy effective field theory we will be based on some phenomenological observations.
Firstly, it is a (2+1)-dimensional phenomenon. The electrons are restricted to the surface of a two-dimensional surface.
Thus the existence of a conserved electric current, $\partial_{\mu}J^{\mu}=0$, enable us to introduce
a vector $A_{\mu}$ as
\begin{equation}
J^{\mu}\sim \epsilon^{\mu\nu\lambda}\partial_{\nu}A_{\lambda},
\label{10.1}
\end{equation}
where $\epsilon^{\mu\nu\lambda}$ is the completely antisymmetric symbol, 
such that the current is automatically conserved. Note that the form (\ref{10.1}) is invariant under the transformation
\begin{equation}
A_{\mu}\rightarrow A_{\mu}+\partial_{\mu}\Lambda,
\label{10.2}
\end{equation}
where $\Lambda$ is an arbitrary function. Thus we are dealing with a truly gauge theory, i.e.,
$A_{\mu}$ is a gauge field. We want a local gauge-invariant Lagrangian
containing only marginal and relevant operators for the gauge field $A_{\mu}$.
We need a bit dimensional analysis to proceed. In the natural system of units, all components of $J^{\mu}$ have
the same dimension, namely, $[J^{\mu}]=2$  in mass units. Consequently, the dimension of the gauge field is $[A_{\mu}]=1$.
The first relevant term that we can imagine is the linear combination $\alpha_{\mu\nu}A^{\mu}A^{\nu}$. It is immediately ruled out by gauge invariance.
Next we can try terms with one derivative. There are several possibilities,
\begin{equation}
A_0\partial_0 A_0,~~~A_0\partial_0 A_i,~~~ A_0\partial_i A_0,~~~A_i\partial_0 A_j,~~~A_0\partial_i A_j,~~~A_i\partial_j A_k,
\label{10.3}
\end{equation}
in addition to other terms that differ from the above ones by total derivatives. All these operators are marginal.
We need to check for gauge invariance.
The only possibility compatible with gauge invariance is the combination $\epsilon^{\mu\nu\sigma}A_{\mu}\partial_{\nu}A_{\sigma}$,
involving the totally antisymmetric symbol $\epsilon^{\mu\nu\sigma}$.
This yields to the Chern-Simons Lagrangian
\begin{equation}
\mathcal{L}=\frac{m}{4\pi}\epsilon^{\mu\nu\sigma}A_{\mu}\partial_{\nu}A_{\sigma}+\cdots,
\label{10.4}
\end{equation}
where $m$ can be identified as the inverse of the filling fraction, $m=1/\nu$.
The dots represent irrelevant operators.
The Chern-Simons is still consistent with the fact that an external magnetic field breaks both parity and time reversal.
Thus the resulting low-energy theory is a Lorentz invariant one. This is an accidental symmetry since it is not
one of the basic requirements. The gauge invariance in three dimensions implies Lorentz invariance for an operator
involving only a single derivative.
This is not so for a Maxwell-type term $F_{\mu\nu}F^{\mu\nu}$. Although it is an irrelevant operator,
as Lorentz symmetry is not a basic assumption, we could expect to have a more general linear combination like $\alpha(F_{0i})^2+\beta(F_{ij})^2$,
since $F_{0i}$ and $F_{ij}$ are independently gauge invariants.

It is interesting to observe that the gauge field $A_{\mu}$ does not represent any physical quantity of
the system. It is not related to the external magnetic field producing the Hall effect. That information is indeed contained in $m$.
The gauge field $A_{\mu}$ is a kind of an emergent quantity reflecting some low-energy physical properties.
It is remarkable that the Chern-Simons action captures much of the physics of the Quantum Hall system, assigning
both fractional charge and statistics for the low-energy excitations, the fractional filling fraction
and the existence of edge currents when we assume a physical boundary in the system.

As a last comment, it is worthy to mention that the Chern-Simons endow a deep invariance,
the topological invariance. This is so because the Chern-Simons action is independent of the metric of the manifold on which it is defined.
This allowed a deep understanding of many properties of the Quantum Hall effect from the topological point of view.
These topological aspects are ultimately responsible for a new chapter in the development of
condensed matter associated to topological orders \cite{Thouless,Wen1,Wen2}.
For example, more recently it led to the theoretical prediction and experimental realization of the topological insulators \cite{Kane1,Hasan,Ando}.
Like the Quantum Hall state, the topological insulators represent a new quantum state of matter which is characterized
by having a bulk band like an ordinary insulator, but have protected conducting states
on their edges.

%%%%%%%%%%%%%%%%%%%%%%%%%%%%%%%%%%%%%%%%%%%%%%

\section{Quantum Gravity, Discrete Spacetime, and Strings}\label{Discrete}

One of the greatest problems of modern physics is how to construct a satisfactory theory of quantum gravity.
When we try to apply the standard methods of perturbation theory to the gravitational interactions governed by the Einstein-Hilbert action
\begin{equation}
S=\int d^4x \sqrt{-g}\left(\frac{1}{16\pi G}R+\mathcal{L}_{matter}\right),
\label{11.1}
\end{equation}
where $R$ is the Ricci scalar and $G$ is the Newton's constant, 
we end up with a nonrenormalizable theory \cite{tHooft}. It means that some UV divergences cannot
be absorbed into the parameters of the theory. Of course, in view of the discussion in Sec. \ref{EFT}
this is not so bad if we are interested in the low-energy physics.
A lot of useful information can be extracted by considering (\ref{11.1}) as an effective field theory
from the beginning \cite{Donoghue}.

But we still could not be content and desire to understand in a more deep way the high-energy regimes of gravitational interactions,
in particular the physical mechanisms and the degrees of freedom relevant in such scales.
We have different paths to follow.
If we insist that the action (\ref{11.1}) correctly describes the quantum properties of gravitational interactions,
we could imagine that the nonrenormalizability is an awkward effect of the perturbative method itself and once the
theory is treated exactly the divergences would disappear.
From the renormalization group perspective this requires the existence of a nontrivial ultraviolet fixed point,
acting as a kind of asymptotic protection to the theory. This proposal is due to Weinberg and it is known as asymptotic
safety \cite{Weinberg}.

Once again, from the effective theory point of view, the action (\ref{11.1}) is a low-energy manifestation.
At the same time we are lead to imagine that a more complex structure, possibly involving completely new physics, can govern
the short-distance structure of the spacetime.  We want to discuss some proposals for that high-energy physics.
One important hint comes from the problem of the strict localization in the presence of gravitational fields.
The equation of motion of (\ref{11.1}) reads
\begin{equation}
R_{\mu\nu}-\frac{1}{2}g_{\mu\nu}R\propto T_{\mu\nu},
\label{11.2}
\end{equation}
where $R_{\mu\nu}$ is the Ricci tensor and $T_{\mu\nu}$ is the energy-momentum tensor. If we try to localize a particle in the spacetime with precision $\Delta x^{\mu}$,
the Heisenberg's uncertainty principle implies an indeterminacy in the energy-momentum of the order $1/\Delta x^{\mu}$.
As we improve the precision of the measurement, more energy-momentum is transmitted to the region of observation.
According to (\ref{11.2}), large contributions to the energy-momentum produce large gravitational fields.
These fields could become so strong such that they prevent light or any other information to scape from the region of observation,
leading to the conclusion that it is a meaningless try to define the strict localization.
We could then imagine that the short-distance structure of the spacetime has an intrinsic discrete nature,
with minimum pieces, which automatically eliminates the concept of points.
In the sense of emergent properties, it suggests that the own concept of a continuum spacetime
is a low-energy manifestation, i.e., a kind of "hydrodynamic" approximation.
Of course, this will lead a breakdown of some familiar concepts as locality, unitarity and causality. But maybe all these concepts are just low-energy manifestations.
See \cite{Seiberg} for an interesting discussion on emergent spacetime.

A possible way to implement these ideas is through the so-called noncommutative theories \cite{Douglas,Szabo}.
The usual spacetime coordinates are replaced by operators, $x^{\mu}\rightarrow \hat{x}^{\mu}$,
satisfying the relation
\begin{equation}
[\hat{x}^{\mu},\hat{x}^{\nu}]=i\Theta^{\mu\nu}.
\label{11.3}
\end{equation}
where $\Theta^{\mu\nu}$ is a constant antisymmetric matrix in the simplest case. It encodes the information about the short-distance structure of the spacetime.
This immediatly introduces a minimum area in the spacetime,
\begin{equation}
\Delta \hat{x}^{\mu} \Delta \hat{x}^{\nu}\geq  \frac{1}{2}| \Theta^{\mu\nu}|,
\label{11.4}
\end{equation}
where $| \Theta^{\mu\nu}|$ is the determinant of the matrix $\Theta^{\mu\nu}$. The continuum spacetime is recovered in the
limit $\Theta^{\mu\nu}\rightarrow 0$.
According to the Weyl-Moyal correspondence, the effect of noncommuting operators can be simulated in terms of
commuting objects but with the ordinary product replaced by the Moyal product,
\begin{equation}
\varphi_1(x)\star \varphi_2(x)\equiv e^{\frac{i}{2}\Theta^{\mu\nu}\partial_{\mu}^x\partial_{\nu}^y}\varphi_1(x)\varphi_2(y) \Big{|}_{y=x}.
\label{11.5}
\end{equation}
In this case, the basic commutation rule (\ref{11.3}) reads $[{x}^{\mu},{x}^{\nu}]_{\star}\equiv x^{\mu}\star x^{\nu}-
x^{\nu}\star x^{\mu}  =i\Theta^{\mu\nu}$.
The presence of derivatives of arbitrary order makes the Moyal product highly nonlocal.
This can be more evident if we
rewrite it as
\begin{equation}
\varphi_1(x)\star \varphi_2(x)=\int d^Dy \int \frac{d^Dk}{(2\pi)^D}\varphi_1(x+\frac{1}{2}\Theta\cdot k)\varphi_2(x+y)e^{iky},
\label{11.6}
\end{equation}
where $(\Theta\cdot k)^{\mu}\equiv \Theta^{\mu\nu}k_{\nu}$, since it involves the product of fields at different spacetime points.
Integration of (\ref{11.6}) reveals an important property of the Moyal product,
\begin{equation}
\int d^D x \,\varphi_1(x)\star \varphi_2(x)= \int d^D x \,\varphi_1(x)\varphi_2(x),
\label{11.7}
\end{equation}
which makes it easy to be implemented in field theories. We just need to introduce the Moyal product in the interaction terms.
The quadratic parts will not be affected and it is simple to obtain the Feynman rules.
Noncommutativity shows up as the presence of trigonometric factors
in the vertices, depending on momenta.
\begin{figure}[!h]
\centering
\includegraphics[scale=0.8]{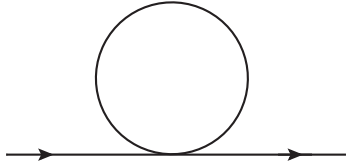}
\caption{One-loop contribution to the two-point function in  $\varphi_{\star}^4$ theory.}
\label{TP}
\end{figure}
We will not go on gravitational theories. It is more instructive to illustrate some basic effects
of noncommutativity in simpler field theories. For example, in a scalar theory with an interaction $\varphi_{\star}^4$, the one-loop
contribution to the two-point function shown in Fig. \ref{TP} is
\begin{equation}
\int \frac{d^4k}{(2\pi)^4}\frac{\cos(\Theta_{\mu\nu}k^{\mu}p^{\nu})}{k^2-m^2}.
\label{11.8}
\end{equation}
The oscillatory factor makes the integral convergent in the UV, but now depending on the external momentum.
For small momenta, it behaves as $1/(\Theta\cdot p)^2$. When inserted into larger diagrams it will
potentially lead to IR divergences that can compromise the renormalizability of the theory.
This is the so-called UV/IR mixing \cite{Seiberg1}. The physics of this type of theory seems to be 
irreconcilable with the decoupling of scales discussed in Sec. \ref{Int}. In general, the usual insensivity of the IR regime to the UV does not hold in noncommutative theories \cite{Pietroni}.

%%%%%%%%%%%%%%%%%%%%%%%%%%%%%%%%%%%%%%%%%%%%

At first sight, the assumption of a discrete microscopic spacetime is tempting since it immediately evades
the severe singularities coming from interactions at coincident points. However, as we discussed,
it leads to a breakdown of some basic notions as locality and unitarity. Another way to alleviate the problems with divergences,
while preserving the continuum character of spacetime, is
by considering the fundamental entities as extended objects like strings, membranes etc,
instead of point-like particles. Thus the interactions between such objects take place at {\it finite regions} of spacetime rather than at coincident points.
It gives the idea that the divergences can become smoother.

Strings are somewhat special among the possibilities \cite{Green,Polchinski3}. This is so because higher dimensional objects than strings possess
so many internal degrees of freedom, corresponding to different configurations they can assume.
In a quantized theory, where it is necessary to sum over all possible configurations,
these many degrees of freedom can yield to internal divergences.
The strings are the only extended objects where both spacetime and internal divergences are
under control \cite{Polchinski3}.
In this picture, the particles are interpreted as different vibrations of a string.
The higher is the excited state bigger is the mass of the corresponding particle.
At low energies we expect that the length of the strings can be neglected recovering the usual particle description.
While the string physical picture is simple, the consequences are rather surprising.
It reveals new ingredients for the microscopic structure of spacetime.
String theory has come a long way until to be recognized as a potential
candidate for a theory involving quantum gravity in addition other interactions but
nowadays it is the most popular candidate to describe the physics at extreme high-energy regimes in a unified and fundamental way.

A basic feature of string description is that the spacetime dimension $D$  where the strings live
is not arbitrary. It is fixed by internal consistency of the theory. In fact, it shows up in a number of ways, but essentially
it is related to the requirement that some of classical symmetries remain true in the quantized theory. 
For the bosonic string $D=26$  while for superstring $D=10$.
In any of the cases, the point is that the theory shows that the microscopic spacetime
dimension can be different from our four dimensional spacetime at low energies.
It leads to the idea that the own dimension is an effective concept.
Once it is assumed that the own spacetime and its dimensionality
can be viewed as emergent concepts, we have more possible theoretical aspects to consider, enlarging
the present framework and feeding the source of ideas to approach this formidable problem.

%%%%%%%%%%%%%%%%%%%%%%%%%%%%%%%%%%%%%%%%%%%%%%%%
\section{Final Considerations}\label{FC}

Symmetries are among the most important concepts of physics and constitute an
indispensable ingredient in the construction and development of all physical theories.
We can separate them in two big groups. A symmetry is said fundamental when it
holds true in the whole spectrum of the system whereas emergent symmetries appear only in specific sectors after some coarse-graining.
Along these notes we have collected several examples which exhibit emergent symmetries.
This phenomenon is ubiquitous in the nature and constitutes a viable mechanism to the realization of symmetries.
The main goal is to illustrate it from different perspectives and discuss some features that can be used in more general contexts.
The key point is that we can take advantage of the emergent symmetries to acquire physical information even in very complex problems.
Effective field theories and the renormalization group are powerful methods to describe the physics and the underlying symmetries in determined scale.

Given a field theory respecting a set of symmetries, we can construct an akin theory violating one or more of them,
which is built up from deformations of the original model. The deformations can be soft or hard, if they involve modifications
of the parameters or of the operator content, respectively. They can also occur simultaneously, where we say that the breaking is
soft+hard. The idea is that if a symmetry is emergent, it is natural looking for theories with deviations of it.
The challenge is to recover it in specific limits of more general theories.
Once we no longer have the constraints from symmetries we have more freedom to write down Lagrangians.
We studied examples of Lorentz, gauge and supersymmetric violating theories,
but the idea can be straightforwardly extended to other symmetries.

Special attention was given to the case of scale and conformal invariance since they are
paradigmatic examples of emergent symmetries at critical points.
They are responsible for the most striking features of critical phenomena, such as the universality.
In two dimensions, the scale invariance is enhanced by conformal invariance which turns out to be
extremely powerful in providing physical information about the critical behavior.

We discussed also some surprising emergent symmetries referred as accidental.
They appear as result of coincident arrangement of the degrees of freedom or properties of the system.
The fact is that this realization of symmetries usually reveals interesting aspects.
For example, while there is an intensive experimental search for supersymmetry in the accelerators, unsuccessfully so far, 
it can be realized in many-body problems of condensed matter and statistical physics \cite{Lee,Yao}.

An important accidental symmetry that was not discussed in the body of the manuscript is the Lorentz
invariance arising in the continuum limit of certain spin models. For example, the effective action obtained for
the two-dimensional Ising model, Eq. (\ref{7.18}), can be written in a manifestly relativistic form
as $i\int d^2x \bar\psi \gamma^{\mu}\partial_{\mu}\psi$. This is a common feature of some spin systems
involving interactions between first-neighbors. A classical example is an antiferromagnetic spin chain, where the
low-energy excitations, with the dispersion relation like $\omega^2\sim k^2$, are described by a relativistic nonlinear sigma model
(see, for example, \cite{Polyakov}).
Other example is the continuum limit of quantum spherical spins models, given by the large-$N$ limit of relativistic nonlinear
sigma models \cite{Vojta,Gomes1}.

%%%%%%%%%%%%%%%%%%%%%%%%%%%%%%%%%%%%%%%%%

The main point we want to emphasize is that the idea of emergent symmetries can be useful to explore deep questions, such as
concerning the microscopic structure of the spacetime and its constituents.
To give an example in this direction, we mention the so-called string-net condensation theory due to Wen \cite{Wen2}.
It gathers some features from spin models, lattice gauge theories and superstrings.
The rich interplay of subjects yield to some interesting physics. In particular,
particles which are usually believed to be fundamental, as electrons and photons, can emerge from the string-net picture as excitations
above the ground state corresponding to certain quantum orders in local bosonic models.

Finally, the study of emergent symmetries, and more generally emergent phenomena,
is a good opportunity to bring together different parts of physics, which provides a fruitful exchange. 
Techniques and concepts coming from one area can find applications in different context generating a confluence of ideas.
With a consistent combination of ingredients, greater are the chances to reach a more accurate description of the nature.

%%%%%%%%%%%%%%%%%%%%%%%%%%%%%%%%%%%%%%%%%%%%%%
\section{Acknowledgments}

It is a pleasure to thank Leandro Bevilaqua, Paula Bienzobaz, Marcelo Gomes, Renann Jusinskas, Silvio Salinas, and Antonio Scarpelli,
for carefully reading the manuscript, useful comments, and many suggestions.
This work was supported by Funda\c{c}\~ao de Amparo a Pesquisa do Estado de S\~ao Paulo (FAPESP).

%%%%%%%%%%%%%%%%%%%%%%%%%%%%%%%%%%%%%%%%%%%%%%%%%%%%%%%

%%%%%%%%%%%%%%%%%%%%%%%%%%%%%%%%%%%%%%%%%%%%
\end{document}